\documentclass{nic-series}
\usepackage{float}
\begin{document} 

\title{Different Kinds of Protein Folding Identified with a Coarse-Grained Heteropolymer Model}

\author{Stefan Schnabel, Michael Bachmann, \and Wolfhard Janke}

\institute{Institut f\"ur Theoretische Physik and Centre for Theoretical Sciences (NTZ),\\ Universit\"at Leipzig, Postfach 100 920, D-04009 Leipzig, Germany\\
         \email{\{Stefan.Schnabel, Michael.Bachmann, Wolfhard.Janke\}@itp.uni-leipzig.de}
          }

\maketitle

\begin{abstracts}
Applying multicanonical simulations we investigated folding properties of off-lattice heteropolymers employing a mesoscopic hydrophobic-polar model. We study for various sequences folding channels in the free-energy landscape by comparing the equilibrium conformations with the folded state in terms of an angular overlap parameter. Although all investigated heteropolymer sequences contain the same content of 
hydrophobic and polar monomers, our analysis of the folding channels reveals a variety of characteristic folding behaviors known from realistic peptides.
\end{abstracts}

\section{Introduction}
The identification of folding channels is one of the key tasks of protein folding studies. While secondary structures depend on atomistic details (such as, in particular, hydrogen bonding), tertiary structure formation should exhibit a certain degree of universality. This suggests that coarse-grained models might capture the main characteristics on mesoscopic scales of this stage of the folding process. In this note we report a computer simulation study that tests this idea by employing the off-lattice hydrophobic-polar AB model.

\section{Model and Method}
In the AB model \cite{AB0} a heteropolymer or coarse-grained peptide is described as a chain of hydrophobic (A) and hydrophilic (B) monomers whose energy is obtained from specific Lennard-Jones potentials between all nonbonded pairs. Thereby an aqueous environment is modeled implicitly by the energetically favored A-A contacts that lead to the formation of a hydrophobic core at low temperatures. A smaller energy contribution arises from the local bending of the chain.

In the Monte Carlo simulations we applied the multicanonical technique \cite{muca} where an additional weight function leads to a flat distribution in energy space. This allows the reweighting of the data to any desired temperature with equally high accuracy. In order to identify folding channels we used the angular overlap parameter $Q$ introduced in Ref.~\citen{handan} to compare equilibrium conformations at temperature $T$ with the previously determined ground-state structure. Two conformations are equal if $Q=1$. The temperature-dependent probability distributions of the total energy $E$ and $Q$, $P_T(E,Q)$, then allow the analysis of the chain's folding behavior (for more details, see Ref. \citen{us}).

\section{Results}
\paragraph{Folding through intermediates:}
For the sequence A\raisebox{-0.5ex}{\scriptsize4}BA\raisebox{-0.5ex}{\scriptsize2}BABA\raisebox{-0.5ex}{\scriptsize2}B\raisebox{-0.5ex}{\scriptsize2}A\raisebox{-0.5ex}{\scriptsize3}BA\raisebox{-0.5ex}{\scriptsize2} the conformations at higher temperatures do not exhibit significant similarities ($Q\approx0.7$) with the ground-state conformation depicted in Fig.~1. As the distributions $P_T(E,Q)$ in Fig.~2 show, the main branch slightly moves to higher $Q$-values with decreasing temperature until it splits below $T\approx0.2$. Near $T\approx0.1$, the population of intermediate conformations has increased and coexists with denatured states. Approaching $T\approx0.05$, the intermediate states dominate the canonical ensemble. The probability for denatured conformations is reduced, but the onset of an occupation of states with similarities to the global-energy minimum (GEM) is clearly visible. At $T=0.02$, the majority of conformations has a large overlap with the ground state and the heteropolymer has folded into its native state.
\begin{figure}[H]
\begin{center}
	      \includegraphics[width=0.2\textwidth]{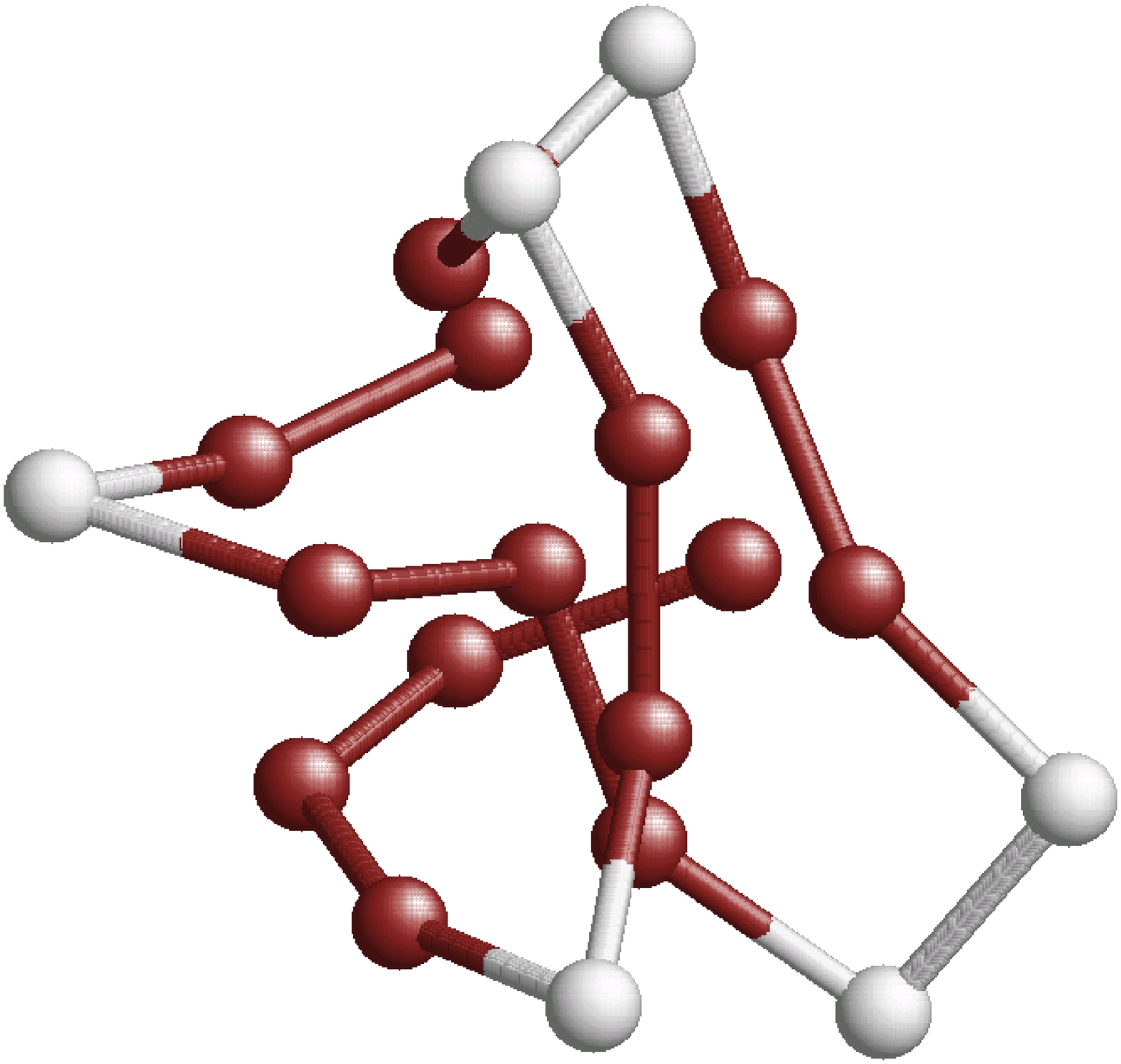}
         \caption[]{\label{20_4} Ground-state conformation of A\raisebox{-0.5ex}{\scriptsize4}BA\raisebox{-0.5ex}{\scriptsize2}BABA\raisebox{-0.5ex}{\scriptsize2}B\raisebox{-0.5ex}{\scriptsize2}A\raisebox{-0.5ex}{\scriptsize3}BA\raisebox{-0.5ex}{\scriptsize2}.}
\end{center}
\end{figure}
\begin{figure}[H]
	\begin{center}
	      \includegraphics[width=0.45\textwidth]{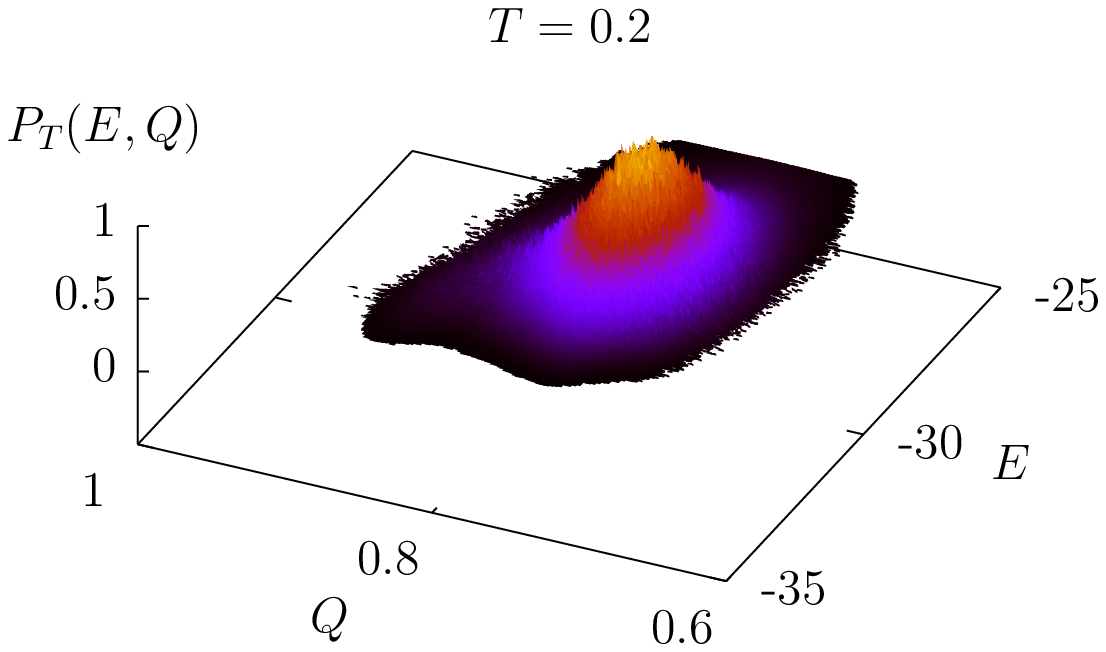}
	      \includegraphics[width=0.45\textwidth]{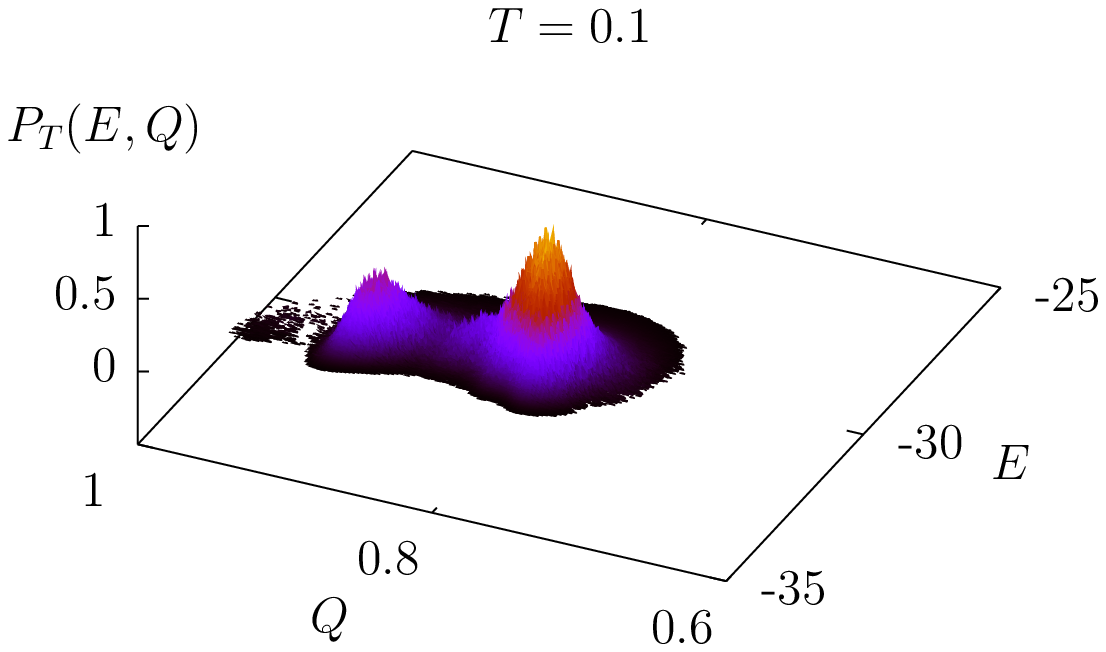}
	      \includegraphics[width=0.45\textwidth]{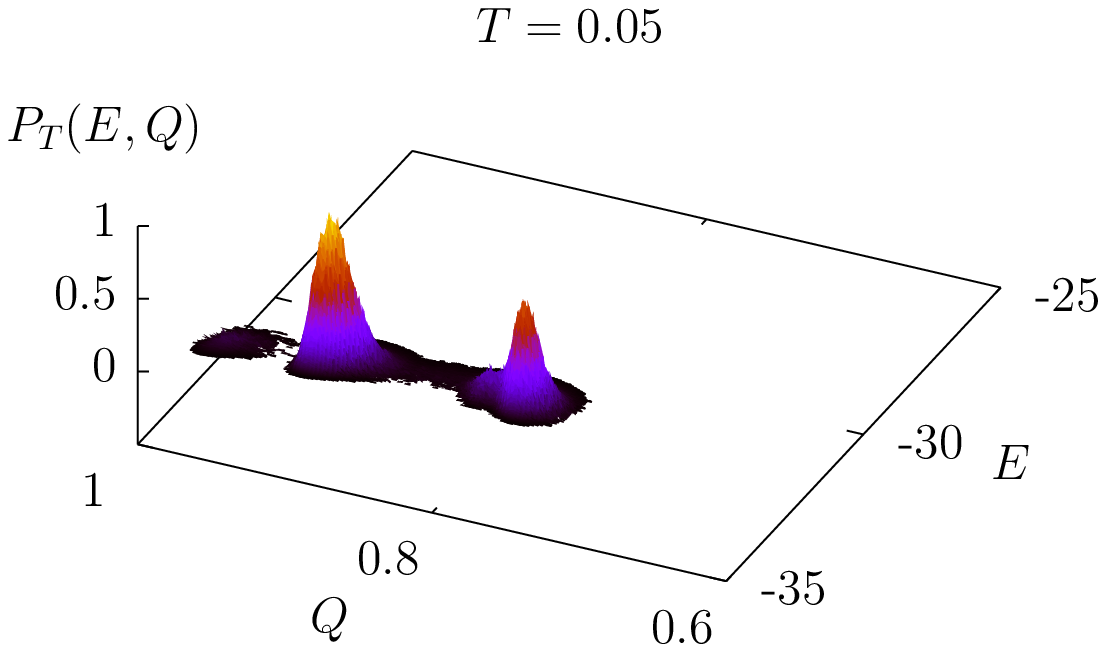}
	      \includegraphics[width=0.45\textwidth]{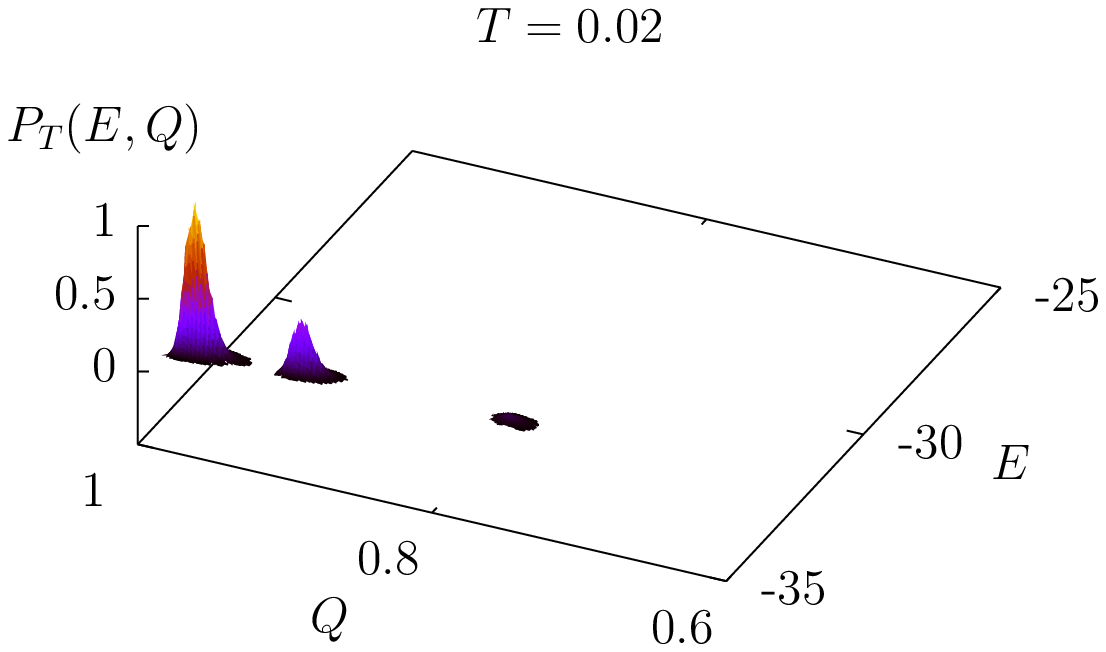}
         \caption[]{\label{S2} Canonical distributions $P_T(E,Q)$ of energy $E$ and overlap $Q$ with the GEM for the sequence A\raisebox{-0.5ex}{\scriptsize4}BA\raisebox{-0.5ex}{\scriptsize2}BABA\raisebox{-0.5ex}{\scriptsize2}B\raisebox{-0.5ex}{\scriptsize2}A\raisebox{-0.5ex}{\scriptsize3}BA\raisebox{-0.5ex}{\scriptsize2}.}
	\end{center}
\end{figure}
\paragraph{Two-state folding and metastability:}
A typical two-state folding scenario can be observed for the sequence BA\raisebox{-0.5ex}{\scriptsize6}BA\raisebox{-0.5ex}{\scriptsize4}BA\raisebox{-0.5ex}{\scriptsize2}BA\raisebox{-0.5ex}{\scriptsize2}B\raisebox{-0.5ex}{\scriptsize2}, where again $Q\approx0.7$ at high temperatures indicates that the heteropolymer is in a random state that possesses no similarities with the GEM conformation. Around $T=0.1$ the canonical ensemble divides and a state with $Q>0.9$ is occupied in addition to the disordered conformations with $Q<0.8$. With decreasing temperature the newly formed branch dominates the ensemble and approaches $Q=1.0$ when $T\rightarrow0$.

An example for metastability is provided by a third heteropolymer with the sequence A\raisebox{-0.5ex}{\scriptsize4}B\raisebox{-0.5ex}{\scriptsize2}A\raisebox{-0.5ex}{\scriptsize4}BA\raisebox{-0.5ex}{\scriptsize2}BA\raisebox{-0.5ex}{\scriptsize3}B\raisebox{-0.5ex}{\scriptsize2}, which exhibits in the low-temperature phase a glassy behavior. While at high temperature similar to the other cases all conformations are distributed around $Q\approx0.7$, the main folding channel does not lead to a single ground-state conformation. Instead two rival conformations ($Q\approx1$ and $Q\approx 0.75$) can be found also at temperatures below $T=0.01$.

\section{Summary}
Numerical investigations of medium size or large proteins by means of Monte Carlo simulations are very difficult since complex energy functions demand high computational efforts. On the other hand, problems of comparable complexity with equal or higher number of degrees of freedom can be handled using simplified coarse-grained models like the AB model. Although the model is relatively simple and the investigated sequences are only permutations of each other (and were not designed especially), three different kinds of folding could be observed. Since all of them -- folding through intermediates, two-state folding and metastability -- are also known from real peptides, the AB model seems to resemble general characteristics of protein folding.\cite{us} Therefore we believe that further research on this model offers the possibility to gain qualitative insights in tertiary folding where microscopic details are expected to be of less importance.

\section*{Acknowledgments}
This work is supported by the DFG (German Science Foundation) under Grant  
No.\ JA 483/24-1/2, the Graduate School ''BuildMoNa'' and the DFH-UFA PhD College CDFA-02-07. Some simulations were performed on the 
supercomputer JUMP of the John von Neumann Institute for Computing (NIC), Forschungszentrum
J\"ulich under Grant No.\ hlz11.

\end{document}